\documentclass{article} 
\usepackage{iclr2025_conference,times}

\usepackage{amsmath,amsfonts,bm}

\def\eqref#1{equation~\ref{#1}}

\def\1{\bm{1}}

\DeclareMathAlphabet{\mathsfit}{\encodingdefault}{\sfdefault}{m}{sl}
\SetMathAlphabet{\mathsfit}{bold}{\encodingdefault}{\sfdefault}{bx}{n}

\usepackage{hyperref}
\usepackage{url}
\usepackage{graphicx}
\usepackage{booktabs}
\usepackage{wrapfig}
\usepackage{subcaption}
\usepackage[capitalise]{cleveref}
\usepackage{todonotes}
\usepackage{amssymb} 
\usepackage{mathabx}
\usepackage{tikz}
\usepackage{placeins} 
\usepackage{enumitem}

\DeclareRobustCommand{\instructIcon}{%
  \scalebox{0.3}{\tikz\draw[fill=black] (90:0.3) -- (234:0.3) -- (18:0.3) -- (162:0.3) -- (306:0.3) -- cycle;}%
}
\DeclareRobustCommand{\safetyIcon}{%
  \scalebox{0.25}{\tikz\draw[fill=black] (0:0.3) -- (60:0.3) -- (120:0.3) -- (180:0.3) -- (240:0.3) -- (300:0.3) -- cycle;}%
}
\DeclareRobustCommand{\advIconUp}{%
  \scalebox{0.7}{$\blacktriangle$}%
}
\DeclareRobustCommand{\advIconDown}{%
  \scalebox{1.1}{$\blacktriangledown$}%
}
\DeclareRobustCommand{\circuitIcon}{%
  \scalebox{0.7}{$\blacklozenge$}%
}
\DeclareRobustCommand{\capabilityCircle}{%
  \tikz\draw[black,fill=black] (0,0) circle (.4ex);%
}
\DeclareRobustCommand{\capabilityIcon}{%
  \scalebox{0.25}{\tikz\draw[fill=black] (90:0.3) -- (162:0.3) -- (234:0.3) -- (306:0.3) -- (18:0.3) -- cycle;}%
}

\title{Fast Proxies for LLM Robustness Evaluation}

\author{Tim Beyer, Jan Schuchardt, Leo Schwinn, Stephan G\"unnemann \\
Technical University of Munich \& Munich Data Science Institute \\
\texttt{\{tim.beyer,j.schuchardt,l.schwinn,s.guennemann\}@tum.de} 
}

\iclrfinalcopy 
\begin{document}

\maketitle

\vspace{-8pt}

\begin{abstract}
Evaluating the robustness of LLMs to adversarial attacks is crucial for safe deployment, yet current red-teaming methods are often prohibitively expensive. 
We compare the ability of fast proxy metrics to predict the real-world robustness of an LLM against a simulated attacker ensemble.
This allows us to estimate a model's robustness to computationally expensive attacks without requiring runs of the attacks themselves.
Specifically, we consider gradient-descent-based embedding-space attacks, prefilling attacks, and direct prompting.
Even though direct prompting in particular does not achieve high ASR, we find that it and embedding-space attacks can predict attack success rates well, achieving $r_p=0.87$ (linear) and $r_s=0.94$ (Spearman rank) correlations with the full attack ensemble while reducing computational cost by three orders of magnitude.
\end{abstract}

\vspace{-8pt}

\section{Introduction}
\label{submission}
As the capabilities of large language models advance, ensuring their robustness and reliability becomes increasingly critical. 
To this end, frontier models undergo extensive adversarial testing and red-teaming to identify vulnerabilities before deployment \citep{openai_red_teaming_network,dubey2024llama}.

However, state-of-the-art red-teaming methods are computationally expensive, as finding adversarial prompts is a challenging combinatorial optimization problem over discrete natural language. 
Here, model-agnostic approaches require prohibitive computational resources~\citep{zou2023universal, chao2023jailbreaking}, whereas more efficient attack algorithms tend to be model-specific and struggle to transfer across architectures~\citep{liao2024amplegcg}. 
Moreover, reliable red-teaming with strong attacks still demands significant manual effort in tailoring the attack algorithm to a specific model~\citep{andriushchenko2024jailbreaking,li2024llm}. 
As a result, large-scale red teaming approaches require thousands of GPU hours~\citep{samvelyan2024rainbow}, making thorough safety evaluations prohibitively expensive in most research settings.

To address this problem, we propose a scalable alternative: low-cost proxies for real-world threat models. 
These proxies enable LLM robustness evaluation without needing to run highly expensive automated attack suites against the model.
As an example of such an attack suite, we use a "synthetic red-teamer" ensemble comprising six distinct LLM attack methods,  which we evaluate on 33 open-source models across 300 harmful prompts. 
We leverage substantial computational resources and aggregate more than 7M jailbreak attempts. 
The data suggest that model robustness in adversarial settings can be predicted through inexpensive approaches.

Our main contributions are as follows:

\begin{itemize}[nosep,noitemsep]
    \item We investigate whether inexpensive proxies --- including direct prompting, prefilling, and embedding space attacks --- can predict robustness against strong adversarial red teaming.
    \item We demonstrate that robustness can be predicted within model families (e.g., different Llama 3 versions) and across model families (e.g., Llama and Mistral).
    \item Finally, we show that by estimating the most robust model checkpoint during training, proxy attacks can aid adversarial model alignment across different training regimes(e.g., circuit breaking or adversarial training).
\end{itemize}

\vspace{-2.5pt}
\section{Synthetic Red-Teamer}\label{section:synethetic_red_teamer}
To emulate a strong attacker, we create a \emph{synthetic red-teamer} by ensembling six common attack algorithms (listed in \cref{tab:ensemble}). All attacks are run using the recommended hyperparameters  (see also
\newpage
\begin{wraptable}{h}{0.455\textwidth}
    \centering
    \setlength{\tabcolsep}{0pt}  
    \caption{Attacks in synthetic red-teamer ensemble \& how many jailbreak candidates they generate per prompt. }
    \begin{tabular}{lr}
        \toprule
        \textbf{Attack Name} & \textbf{Candidates} \\
        \midrule
        AmpleGCG \citep{liao2024amplegcg} & 200 \\
        AutoDAN \citep{liu2023autodan} & 100 \\
        BEAST \citep{sadasivan2024fast} & 40 \\
        GCG \citep{zou2023universal} & 250 \\
        HumanJB \citep{mazeika2024harmbench} & 112 \\
        PAIR \citep{chao2023jailbreaking} & 25  \\
        \midrule
        Total & 727 \\
        \bottomrule
    \end{tabular}
    \vskip -0.35cm
    \label{tab:ensemble}
\end{wraptable}
 \cref{sec:hyperparameters}) and simulate a strong red-teamer with significant computational resources ($\approx$ 30 H100-minutes per prompt).
We evaluate each algorithm in the \textit{many-trial} setting, where all candidate prompts (including intermediate steps) are tried on the victim model. 
Thus, for each harmful prompt in the dataset, a model is attacked by 727 different input prompts.
If \emph{any} of the prompts succeed, we count the attack as successful.
While some algorithms (e.g., AmpleGCG and PAIR) perform many-trial attacks by default, others, such as GCG and BEAST, generally only use the final attack prompt to generate a harmful response.   
The many-trial setting makes attacks strictly more powerful, at the cost of increased compute.

\section{Proxy Methods}

We aim to find an inexpensive and fast approach that can reliably predict a model's real-world robustness.
Finding such a proxy for robustness could dramatically reduce the cost of robustness evaluations, make it easier to compare models across and within families, and efficiently select promising checkpoints during defense training.
To this end, we consider three candidate approaches:

\textbf{Embedding Space Attacks.} 
\citet{schwinn2023adversarial,schwinn2024soft} recently proposed a white box attack that operates in continuous token embedding space, rather than the discrete input vocabulary. 
This framework---while impractical for real-world attacks, where most threat models assume a black box setting with string-level input---provides an extremely fast way to attack models in a white box setting, and can be used e.g., to adversarially train LLMs \citep{xhonneux2024efficient}.

\textbf{Prefilling.} 
Prefilling attacks \citep{vega2023bypassing,andriushchenko2024jailbreaking} rely on injecting a prefix to the beginning of the victim model's response to the harmful prompt - typically using an affirmative response prefix. 
As this level of access is also provided by some private models (e.g., the Claude family \citep{TheC3}), it represents a realistic attack vector even for hosted models.

\textbf{Direct.} \emph{Direct} prompting is the simplest possible baseline: 
We simply use an unmodified harmful prompt from the dataset and sample a single greedy generation, which is then judged.

\section{Experimental Evaluation}

We conduct experiments to determine how well the attack success rates of inexpensive proxy methods (direct ASR, prefilling ASR, embedding-space ASR) predict robustness against real-world red-teaming approaches, which we simulate using our strong synthetic red-teamer from~\cref{section:synethetic_red_teamer} across various training and attack scenarios.
In addition to directly comparing the different ASRs, we compute Pearson correlation ($r_p$) to quantify linear correlation between proxy ASR and ensemble ASR. 
We further compute Spearman ($r_s)$ and Kendall rank ($\tau$) correlation to understand whether the order of any two models w.r.t.\ proxy ASR is predictive of their order w.r.t.\ ensemble ASR.
For the full details of our experimental setup, see \cref{sec:hyperparameters}. For additional results see \cref{sec:additional-results}.

\subsection{Comparing Within-Family Models}
\begin{figure}[h!]
    \includegraphics[width=\textwidth]{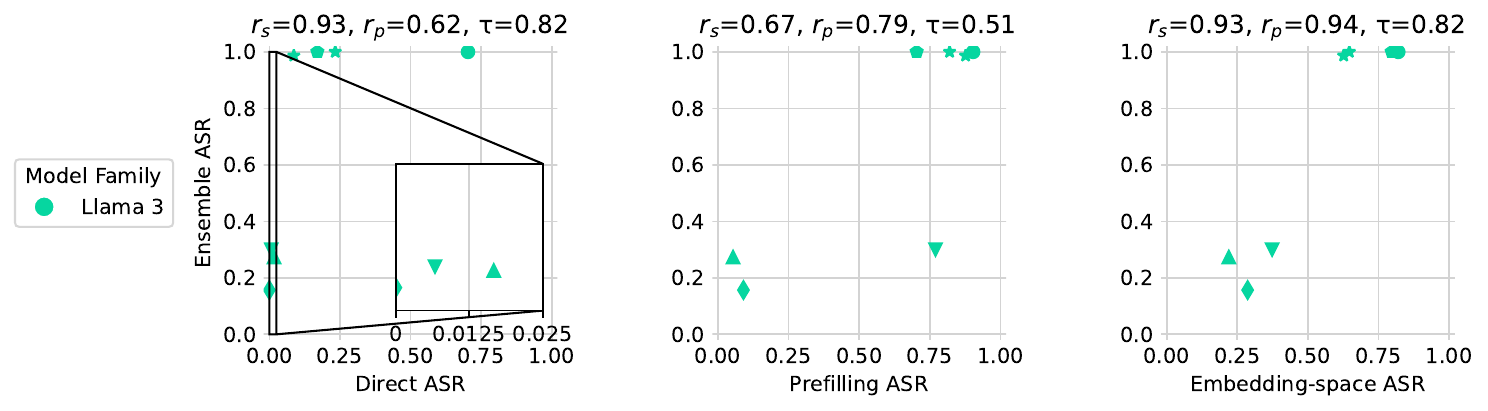}
    \vskip -0.35cm
    \caption{Attack success rates for different variants of Llama-3-8B. We include instruct versions (\instructIcon) as a baseline and compare to safety-tuned (\safetyIcon), adversarially trained (\advIconUp,\advIconDown), circuit breaker (\circuitIcon), and capability-optimized (\capabilityCircle,\capabilityIcon) models.}
    \label{fig:combined_correlations_individual_family_llama3}
    \vskip -0.5cm
\end{figure}
Popular base models are often fine-tuned for particular use cases, such as chatting \citep{tunstall2023zephyr}, helpfulness \citep{starling2023}, or tool use \citep{teknium2024hermes}.
We are interested in comparing the safety of several post-trained model versions.
In Figure \ref{fig:combined_correlations_individual_family_llama3}, we evaluate different derivatives of Llama 3 8B Instruct.
Spearman and Kendall rank correlation coefficients $r_s$ and $\tau$ of direct prompting are greater or equal than those of the other proxy attacks.
We observe that direct ASR is close to $0$ for multiple models, which impedes a good linear fit ($r_p$ of $0.62$) between direct ASR and ensemble ASR. This $r_p$ is smaller than those of prefilling and embedding space attacks. 
Thus, even for within-family comparisons, the simplest and fastest attack appears like a suitable choice as a proxy for computationally expensive red-teaming.

\subsection{Comparing Across Model Families}\label{section:comparing_families}

\begin{figure}[h]
    \centering
    \includegraphics[width=\textwidth]{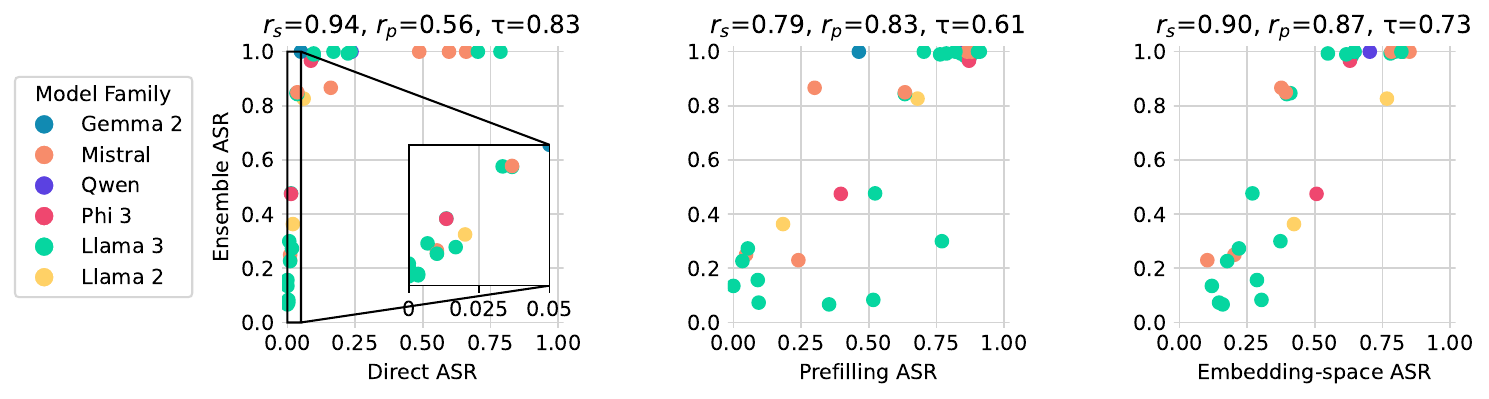}
    \vskip -0.35cm
    \caption{
        Attack success rates for models from different families.
        Direct ASR has the largest $r_s$ and $\tau$, i.e., 
        the order of two models w.r.t. direct ASR is most predictive of order w.r.t. ensemble ASR.
    }
    \label{fig:combined_correlations_family}
\end{figure}

We also investigate whether proxy methods can be used to predict the success rate of expensive red-teaming attacks on newly introduced model families.
In~\cref{fig:combined_correlations_family}, 
each point corresponds to a specific model from one of six model families (Gemma 2 \citep{team2024gemma}, Mistral \citep{jiang2023mistral}, Qwen \citep{bai2023qwen}, Phi-3 \citep{abdin2024phi}, Llama 3 \citep{dubey2024llama}, Llama 2 \citep{touvron2023llama}).
Prefilling and embedding space attacks often have much higher ASR than direct prompting, which na{\"i}vely use the harmful prompt without any modification.
Direct ASR is generally below $5\%$, except for models that are extremely unrobust (ensemble ASR close to $100\%$).
Thus, the pairs of direct and ensemble ASR do not admit a linear fit and the Pearson correlation $r_p$ is small.
However, the rank correlation coefficients of direct prompting ($r_s=0.94$, $\tau=0.83$) are higher than those of the other two proxy methods ($r_s=0.79$, $\tau=0.61$) and ($r_s=0.90$, $\tau=0.73$). 

\subsection{Assessing Effectiveness of Robustness Fine-Tuning}

\begin{figure}[h]
        \includegraphics[width=\textwidth]{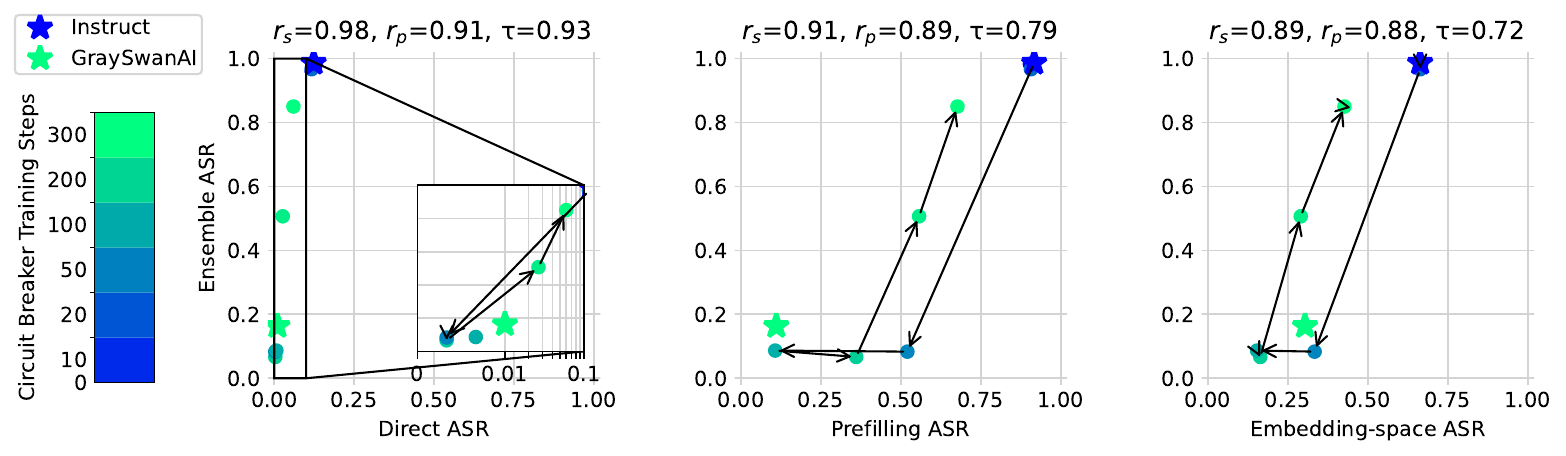}
        \vskip -0.35cm
        \caption{Attack success rates for different number of robustness fine-tuning steps using Circuit Breakers \citep{zou2024improving}. We include the base instruct model and the officially released circuit breaker model.
        Despite varying success rate, all proxy methods have similar correlation coefficients, i.e., are similarly predictive of fine-tuning effectiveness. Arrows indicate training progression.
        }
        \label{fig:combined_correlations_training_cb}
\end{figure}

A standard method for increasing model robustness is via post-training/fine-tuning approaches, e.g., via circuit breaker training~\citep{zou2023universal} or continuous adversarial training~\citep{sheshadri2024targeted, xhonneux2024efficient}.
In~\cref{fig:combined_correlations_training_cb} \& \ref{fig:combined_correlations_training_capo}, we assess whether proxy ASR can potentially be used to predict ensemble ASR after fine-tuning for a specific number of steps, rather than performing computationally expensive red-teaming for every possible value of this hyper-parameter.
Specifically, we apply circuit breaker training to Llama-3-8B-Instruct and vary the number of training steps between $1$ and $300$.
Again, while the relation between proxy ASR and ensemble ASR is generally monotonic and linear for all three proxies, direct prompting achieves significantly higher ranking correlations $r_s$ and $\tau$.

\subsection{Scaling Trends}
\begin{figure}[h]
    \centering
    \includegraphics[width=\textwidth]{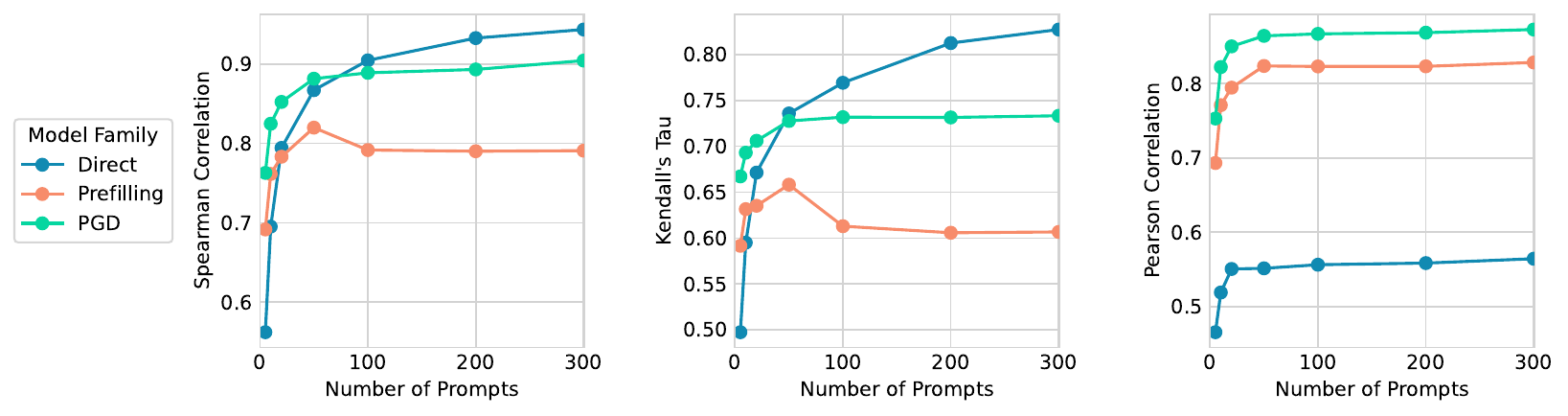}
    \vskip -0.3cm
    \caption{Correlation coefficients between proxy attack success rate and ensemble attack success rate under varying number of prompts. When using fewer than 50 prompts, PGD yields higher Spearman and Kendall ranking correlations, however the direct prompting scales better with more prompts. Prefilling and PGD achieve higher linear/Pearson correlations at any prompt count.}
    \label{fig:scaling_trends}
\end{figure}

We find that the effectiveness of different proxy methods varies with the amount of prompts used (\cref{fig:scaling_trends}).
Prefilling and embedding space attacks attain universally higher Pearson correlation, i.e., admit a better linear fit irrespective of the number of prompts. 
They can also reach higher Spearman and Kendall ranking correlation --- but only when using few prompts.
For $50$ or more prompts, direct prompting yields higher ranking correlation coefficients.
This can be explained as follows:
Since direct ASR is generally small for robust models, there is a high chance that our sample estimate will incorrectly indicate a direct ASR of exactly $0$ when using few prompts, making the observed relation to ensemble ASR very erratic.
Using more prompts provides a better estimate of the small but non-zero population success rate of direct prompting, thus eliminating this issue and making direct ASR a good predictor of whether one model will be more robust than another to our synthetic red-teamer. 
As increasingly robust models will decrease ASR, we expect to see an increase in the number of prompts required to effectively use direct prompting as a proxy.

\subsection{Limitations}
While we conducted an exhaustive and computationally intensive evaluation using six attacks and 33 models from the sub-10B parameter class, our experiments should be further validated to ensure they generalize to other attack algorithms and model sizes.

\section{Conclusion}
We investigated the effectiveness of inexpensive proxy attacks in predicting LLM robustness against adversarial red-teaming.
Our results highlight key trade-offs between different proxy methods.
Direct prompting is a strong baseline for ranking models by robustness across diverse scenarios (within-family, cross-family, safety fine-tuning), provided that enough ($>50$) prompts are used.
Embedding-space attacks provide better ranking at low prompt count and better linear fits, while prefilling attacks are generally inferior to the two alternatives.
Overall, our results showcase that efficient proxy attacks are a promising direction for future research
towards making foundation models more responsible without incurring unjustifiable computational overhead.

\bibliography{iclr2025_conference}

\begin{thebibliography}{26}
\providecommand{\natexlab}[1]{#1}
\providecommand{\url}[1]{\texttt{#1}}
\expandafter\ifx\csname urlstyle\endcsname\relax
  \providecommand{\doi}[1]{doi: #1}\else
  \providecommand{\doi}{doi: \begingroup \urlstyle{rm}\Url}\fi

\bibitem[Abdin et~al.(2024)Abdin, Aneja, Behl, Bubeck, Eldan, Gunasekar, Harrison, Hewett, Javaheripi, Kauffmann, et~al.]{abdin2024phi}
Marah Abdin, Jyoti Aneja, Harkirat Behl, S{\'e}bastien Bubeck, Ronen Eldan, Suriya Gunasekar, Michael Harrison, Russell~J Hewett, Mojan Javaheripi, Piero Kauffmann, et~al.
\newblock Phi-4 technical report.
\newblock \emph{arXiv preprint arXiv:2412.08905}, 2024.

\bibitem[Andriushchenko et~al.(2024)Andriushchenko, Croce, and Flammarion]{andriushchenko2024jailbreaking}
Maksym Andriushchenko, Francesco Croce, and Nicolas Flammarion.
\newblock Jailbreaking leading safety-aligned {LLMs} with simple adaptive attacks.
\newblock \emph{arXiv preprint arXiv:2404.02151}, 2024.

\bibitem[Anthropic(2024)]{TheC3}
Anthropic.
\newblock The {C}laude 3 model family: {O}pus, {S}onnet, {H}aiku.
\newblock 2024.
\newblock URL \url{https://api.semanticscholar.org/CorpusID:268232499}.

\bibitem[Bai et~al.(2023)Bai, Bai, Chu, Cui, Dang, Deng, Fan, Ge, Han, Huang, et~al.]{bai2023qwen}
Jinze Bai, Shuai Bai, Yunfei Chu, Zeyu Cui, Kai Dang, Xiaodong Deng, Yang Fan, Wenbin Ge, Yu~Han, Fei Huang, et~al.
\newblock Qwen technical report.
\newblock \emph{arXiv preprint arXiv:2309.16609}, 2023.

\bibitem[Chao et~al.(2023)Chao, Robey, Dobriban, Hassani, Pappas, and Wong]{chao2023jailbreaking}
Patrick Chao, Alexander Robey, Edgar Dobriban, Hamed Hassani, George~J Pappas, and Eric Wong.
\newblock Jailbreaking black box large language models in twenty queries.
\newblock \emph{arXiv preprint arXiv:2310.08419}, 2023.

\bibitem[Dubey et~al.(2024)Dubey, Jauhri, Pandey, Kadian, Al-Dahle, Letman, Mathur, Schelten, Yang, Fan, et~al.]{dubey2024llama}
Abhimanyu Dubey, Abhinav Jauhri, Abhinav Pandey, Abhishek Kadian, Ahmad Al-Dahle, Aiesha Letman, Akhil Mathur, Alan Schelten, Amy Yang, Angela Fan, et~al.
\newblock The {L}lama 3 herd of models.
\newblock \emph{arXiv preprint arXiv:2407.21783}, 2024.

\bibitem[Jiang et~al.(2023)Jiang, Sablayrolles, Mensch, Bamford, Chaplot, Casas, Bressand, Lengyel, Lample, Saulnier, et~al.]{jiang2023mistral}
Albert~Q Jiang, Alexandre Sablayrolles, Arthur Mensch, Chris Bamford, Devendra~Singh Chaplot, Diego de~las Casas, Florian Bressand, Gianna Lengyel, Guillaume Lample, Lucile Saulnier, et~al.
\newblock Mistral 7{B}.
\newblock \emph{arXiv preprint arXiv:2310.06825}, 2023.

\bibitem[Li et~al.(2024)Li, Han, Steneker, Primack, Goodside, Zhang, Wang, Menghini, and Yue]{li2024llm}
Nathaniel Li, Ziwen Han, Ian Steneker, Willow Primack, Riley Goodside, Hugh Zhang, Zifan Wang, Cristina Menghini, and Summer Yue.
\newblock Llm defenses are not robust to multi-turn human jailbreaks yet.
\newblock \emph{arXiv preprint arXiv:2408.15221}, 2024.

\bibitem[Liao \& Sun(2024)Liao and Sun]{liao2024amplegcg}
Zeyi Liao and Huan Sun.
\newblock Ample{GCG}: {L}earning a universal and transferable generative model of adversarial suffixes for jailbreaking both open and closed llms.
\newblock \emph{arXiv preprint arXiv:2404.07921}, 2024.

\bibitem[Liu et~al.(2023)Liu, Xu, Chen, and Xiao]{liu2023autodan}
Xiaogeng Liu, Nan Xu, Muhao Chen, and Chaowei Xiao.
\newblock Autodan: {G}enerating stealthy jailbreak prompts on aligned large language models.
\newblock \emph{arXiv preprint arXiv:2310.04451}, 2023.

\bibitem[Mazeika et~al.(2024)Mazeika, Phan, Yin, Zou, Wang, Mu, Sakhaee, Li, Basart, Li, et~al.]{mazeika2024harmbench}
Mantas Mazeika, Long Phan, Xuwang Yin, Andy Zou, Zifan Wang, Norman Mu, Elham Sakhaee, Nathaniel Li, Steven Basart, Bo~Li, et~al.
\newblock Harm{B}ench: {A} standardized evaluation framework for automated red teaming and robust refusal.
\newblock \emph{arXiv preprint arXiv:2402.04249}, 2024.

\bibitem[OpenAI(2023)]{openai_red_teaming_network}
OpenAI.
\newblock Openai red teaming network, 2023.
\newblock URL \url{https://openai.com/index/red-teaming-network/}.

\bibitem[Sadasivan et~al.(2024)Sadasivan, Saha, Sriramanan, Kattakinda, Chegini, and Feizi]{sadasivan2024fast}
Vinu~Sankar Sadasivan, Shoumik Saha, Gaurang Sriramanan, Priyatham Kattakinda, Atoosa Chegini, and Soheil Feizi.
\newblock Fast adversarial attacks on language models in one {GPU} minute.
\newblock \emph{arXiv preprint arXiv:2402.15570}, 2024.

\bibitem[Samvelyan et~al.(2024)Samvelyan, Raparthy, Lupu, Hambro, Markosyan, Bhatt, Mao, Jiang, Parker-Holder, Foerster, et~al.]{samvelyan2024rainbow}
Mikayel Samvelyan, Sharath~Chandra Raparthy, Andrei Lupu, Eric Hambro, Aram~H Markosyan, Manish Bhatt, Yuning Mao, Minqi Jiang, Jack Parker-Holder, Jakob Foerster, et~al.
\newblock Rainbow teaming: Open-ended generation of diverse adversarial prompts.
\newblock \emph{arXiv preprint arXiv:2402.16822}, 2024.

\bibitem[Schwinn et~al.(2023)Schwinn, Dobre, G{\"u}nnemann, and Gidel]{schwinn2023adversarial}
Leo Schwinn, David Dobre, Stephan G{\"u}nnemann, and Gauthier Gidel.
\newblock Adversarial attacks and defenses in large language models: Old and new threats.
\newblock \emph{arXiv preprint arXiv:2310.19737}, 2023.

\bibitem[Schwinn et~al.(2024)Schwinn, Dobre, Xhonneux, Gidel, and Gunnemann]{schwinn2024soft}
Leo Schwinn, David Dobre, Sophie Xhonneux, Gauthier Gidel, and Stephan Gunnemann.
\newblock Soft prompt threats: {A}ttacking safety alignment and unlearning in open-source {LLMs} through the embedding space.
\newblock \emph{arXiv preprint arXiv:2402.09063}, 2024.

\bibitem[Sheshadri et~al.(2024)Sheshadri, Ewart, Guo, Lynch, Wu, Hebbar, Sleight, Stickland, Perez, Hadfield-Menell, and Casper]{sheshadri2024targeted}
Abhay Sheshadri, Aidan Ewart, Phillip Guo, Aengus Lynch, Cindy Wu, Vivek Hebbar, Henry Sleight, Asa~Cooper Stickland, Ethan Perez, Dylan Hadfield-Menell, and Stephen Casper.
\newblock Targeted latent adversarial training improves robustness to persistent harmful behaviors in llms.
\newblock \emph{arXiv preprint arXiv:2407.15549}, 2024.

\bibitem[Team et~al.(2024)Team, Riviere, Pathak, Sessa, Hardin, Bhupatiraju, Hussenot, Mesnard, Shahriari, Ram{\'e}, et~al.]{team2024gemma}
Gemma Team, Morgane Riviere, Shreya Pathak, Pier~Giuseppe Sessa, Cassidy Hardin, Surya Bhupatiraju, L{\'e}onard Hussenot, Thomas Mesnard, Bobak Shahriari, Alexandre Ram{\'e}, et~al.
\newblock Gemma 2: Improving open language models at a practical size.
\newblock \emph{arXiv preprint arXiv:2408.00118}, 2024.

\bibitem[Teknium et~al.(2024)Teknium, Quesnelle, and Guang]{teknium2024hermes}
Ryan Teknium, Jeffrey Quesnelle, and Chen Guang.
\newblock Hermes 3 technical report.
\newblock \emph{arXiv preprint arXiv:2408.11857}, 2024.

\bibitem[Touvron et~al.(2023)Touvron, Martin, Stone, Albert, Almahairi, Babaei, Bashlykov, Batra, Bhargava, Bhosale, et~al.]{touvron2023llama}
Hugo Touvron, Louis Martin, Kevin Stone, Peter Albert, Amjad Almahairi, Yasmine Babaei, Nikolay Bashlykov, Soumya Batra, Prajjwal Bhargava, Shruti Bhosale, et~al.
\newblock Llama 2: {O}pen foundation and fine-tuned chat models.
\newblock \emph{arXiv preprint arXiv:2307.09288}, 2023.

\bibitem[Tunstall et~al.(2023)Tunstall, Beeching, Lambert, Rajani, Rasul, Belkada, Huang, von Werra, Fourrier, Habib, et~al.]{tunstall2023zephyr}
Lewis Tunstall, Edward Beeching, Nathan Lambert, Nazneen Rajani, Kashif Rasul, Younes Belkada, Shengyi Huang, Leandro von Werra, Cl{\'e}mentine Fourrier, Nathan Habib, et~al.
\newblock Zephyr: Direct distillation of lm alignment.
\newblock \emph{arXiv preprint arXiv:2310.16944}, 2023.

\bibitem[Vega et~al.(2023)Vega, Chaudhary, Xu, and Singh]{vega2023bypassing}
Jason Vega, Isha Chaudhary, Changming Xu, and Gagandeep Singh.
\newblock Bypassing the safety training of open-source llms with priming attacks.
\newblock \emph{arXiv preprint arXiv:2312.12321}, 2023.

\bibitem[Xhonneux et~al.(2024)Xhonneux, Sordoni, G{\"u}nnemann, Gidel, and Schwinn]{xhonneux2024efficient}
Sophie Xhonneux, Alessandro Sordoni, Stephan G{\"u}nnemann, Gauthier Gidel, and Leo Schwinn.
\newblock Efficient adversarial training in {LLMs} with continuous attacks.
\newblock \emph{arXiv preprint arXiv:2405.15589}, 2024.

\bibitem[Zhu et~al.(2023)Zhu, Frick, Wu, Zhu, and Jiao]{starling2023}
Banghua Zhu, Evan Frick, Tianhao Wu, Hanlin Zhu, and Jiantao Jiao.
\newblock Starling-7b: Improving llm helpfulness \& harmlessness with rlaif, November 2023.

\bibitem[Zou et~al.(2023)Zou, Wang, Carlini, Nasr, Kolter, and Fredrikson]{zou2023universal}
Andy Zou, Zifan Wang, Nicholas Carlini, Milad Nasr, J~Zico Kolter, and Matt Fredrikson.
\newblock Universal and transferable adversarial attacks on aligned language models.
\newblock \emph{arXiv preprint arXiv:2307.15043}, 2023.

\bibitem[Zou et~al.(2024)Zou, Phan, Wang, Duenas, Lin, Andriushchenko, Kolter, Fredrikson, and Hendrycks]{zou2024improving}
Andy Zou, Long Phan, Justin Wang, Derek Duenas, Maxwell Lin, Maksym Andriushchenko, J~Zico Kolter, Matt Fredrikson, and Dan Hendrycks.
\newblock Improving alignment and robustness with circuit breakers.
\newblock In \emph{The Thirty-eighth Annual Conference on Neural Information Processing Systems}, 2024.

\end{thebibliography}
\bibliographystyle{iclr2025_conference}

\appendix
\newpage
\section{Model Zoo}
\label{sec:model-zoo}

\begin{table}[h!]
    \centering
    \begin{tabular}{ll}
    \toprule
        \textbf{HuggingFace Model ID} & \textbf{Model Family} \\
        \midrule
        google/gemma-2-2b-it & Gemma 2 \\
        mistralai/Mistral-7B-Instruct-v0.3 & Mistral 7B \\
        berkeley-nest/Starling-LM-7B-alpha & Mistral 7B \\
        cais/zephyr\_7b\_r2d2 & Mistral 7B \\
        HuggingFaceH4/zephyr-7b-beta & Mistral 7B \\
        ContinuousAT/Zephyr-CAT & Mistral 7B \\
        GraySwanAI/Mistral-7B-Instruct-RR & Mistral 7B \\
        mistralai/Mistral-Nemo-Instruct-2407 & Mistral Nemo \\
        mistralai/Ministral-8B-Instruct-2410 & Ministral \\
        meta-llama/Llama-2-7b-chat-hf & Llama 2 \\
        ContinuousAT/Llama-2-7B-CAT & Llama 2 \\
        lmsys/vicuna-7b-v1.5 & Llama 2 \\
        NousResearch/Hermes-2-Pro-Llama-3-8B & Llama 3 \\
        meta-llama/Meta-Llama-3-8B-Instruct & Llama 3 \\
        LLM-LAT/robust-llama3-8b-instruct & Llama 3 \\
        GraySwanAI/Llama-3-8B-Instruct-RR & Llama 3 \\
        meta-llama/Meta-Llama-3.1-8B-Instruct & Llama 3.1 \\
        allenai/Llama-3.1-Tulu-3-8B-DPO & Llama 3.1 \\
        meta-llama/Llama-3.2-1B-Instruct & Llama 3.2 \\
        meta-llama/Llama-3.2-3B-Instruct & Llama 3.2 \\
        qwen/Qwen2-7B-Instruct & Qwen2 7B \\
        ContinuousAT/Phi-CAT & Phi 3 \\
        microsoft/Phi-3-mini-4k-instruct & Phi 3 mini \\
        microsoft/phi-4 & Phi 4 \\
        \bottomrule
    \end{tabular}
    \caption{List of models with their short names and base models.}
    \label{tab:models}
\end{table}

In addition, we fine-tune Llama-3-8B-Instruct using the circuit breaker methodology \citep{zou2024improving} using $N=\{1,10,20,50,100,200,300,500,1000\}$ steps, and with the CAPO version of continuous adversarial training \citep{xhonneux2024efficient} and $N=\{75,150,225\}$ steps.
We use \texttt{bfloat16} quantization for all models.

\newpage
\section{Hyperparameters \& Experimental Details}
\label{sec:hyperparameters}
We run all attacks on all 300 harmful prompts from AdvBench \citep{zou2023universal}, as included in HarmBench. 
A jailbreak attempt is counted as successful if both HarmBench's finetuned Llama-2-13B classifier \citep{mazeika2024harmbench} and LlamaGuard 3 8B \citep{dubey2024llama} flag the model's response as harmful.

The hyperparameters for the attacks used in the ensemble and the proxy attacks are shown below. 
Where possible, attack implementations were sourced from the original authors’ GitHub repositories; otherwise, we integrated a HarmBench implementation into our pipeline. 
In some cases, we consulted authors directly to obtain reference implementations and verify correctness. For all attacks, we evaluate a single greedy generation per prompt-candidate.

\begin{itemize}
    \item AmpleGCG \cite{liao2024amplegcg}: We use \texttt{osunlp/AmpleGCG-llama2-sourced-} \texttt{llama2-7b-chat} to generate 200 attack suffixes with diversity penalty 1 and generate completions for all 200 of the attack candidates. 
    \item AutoDAN \cite{liu2023autodan}: We use 100 steps and initialize using the 128 seed prompts from HarmBench's implementation. We use the attacked model itself as mutator model and set $N_\text{elites}=0.05$, $\text{crossover}=0.5$, $N_\text{points}=5$, and $P_\text{mutation}=0.01$.
    \item BEAST \cite{sadasivan2024fast}: We use $k1=k2=15$ and set the temperature to 1 to sample $N=40$  suffix tokens.
    \item GCG \cite{zou2023universal}: We use a modified version of nanoGCG with a corrected token filtering algorithm to remove special tokens \& only allow ASCII-representable characters. We set $N_\text{steps}=250$, use a batch size of 512, $\text{Top-K}=256$, and initialize using the string ``x x x x x x x x x x x x x x x x x x x x" as it tokenizes to exactly 20 tokens for all tested models.
    \item HumanJailbreaks: We use the \href{https://github.com/centerforaisafety/HarmBench/blob/main/baselines/human_jailbreaks/jailbreaks.py}{114 human-designed jailbreak templates} in HarmBench \cite{mazeika2024harmbench} to prompt the model.
    \item PAIR \cite{chao2023jailbreaking}: We use \texttt{lmsys/vicuna-13b-v1.5} as attacker model and generate up to 512 tokens per attack prompt. Sampling attacks is done with with temperature 1 and top-p of 0.9, setting $N_\text{streams}$ to 5 and $N_\text{iterations}$ to 5. During the attack, the victim model generates up to 256 tokens using greedy generation. If the conversations grow longer than the model's context, we truncate the first non-system messages from the conversation until the conversation fits into the context window.
    \end{itemize}
        
The proxy attacks use the following settings: 
\begin{itemize}
    \item Direct: We simply use the harmful prompt without any modification and sample a greedy generation.
    \item Embedding-space  \cite{schwinn2024soft}: We initialize the attack using the suffix ``x x x x x x x x x x x x x x x x x x x x" as it tokenizes to exactly 20 tokens for all tested models, and run signed gradient descent optimization for 100 steps. We use a learning rate of $\alpha=0.01$ and constrain the optimization to an $L_2$ ball with radius 1 around the initialization for each token. To normalize across model families, we normalize both step size and $L_2$ constaint by the average $L_2$ embedding norm across the input vocabulary.
    \item Prefilling: We use the unmodified harmful prompt and pre-fill the beginning of the model's response using the affirmative target sequence from the dataset.
\end{itemize}

Running the attack ensemble on an Nvidia H100 GPU for a single prompt requires 1,731 seconds on average, while direct prompting and prefilling can be easily batched and is completed in a single second.
Batched embedding space attacks require approximately 5 seconds per prompt.

\newpage
\section{Additional Experimental Results}
\label{sec:additional-results}
\subsection{Mistral Variants}

\begin{figure}[h]
    \includegraphics[width=\textwidth]{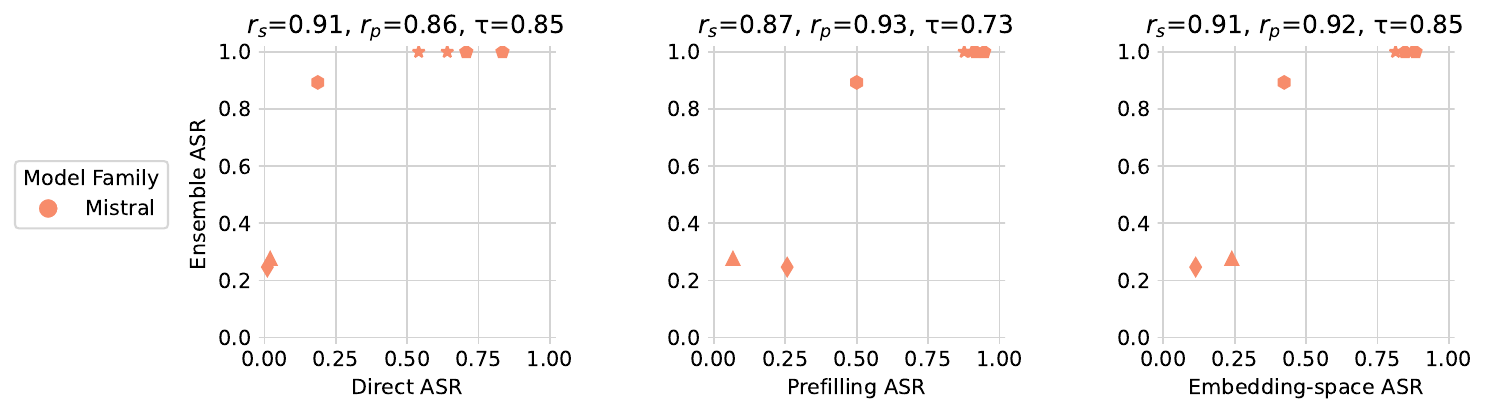}
    \caption{Attack success rates for different variants of Mistral 7B Instruct. We include instruct versions (\instructIcon) as a baseline and compare to safety-tuned (\safetyIcon), adversarially trained (\advIconUp), circuit breaker (\circuitIcon), and capability-optimized (\capabilityIcon) models.}
    \label{fig:combined_correlations_individual_family_mistral}
\end{figure}

\subsection{Llama 3 Variants}

\begin{figure}[h]
    \includegraphics[width=\textwidth]{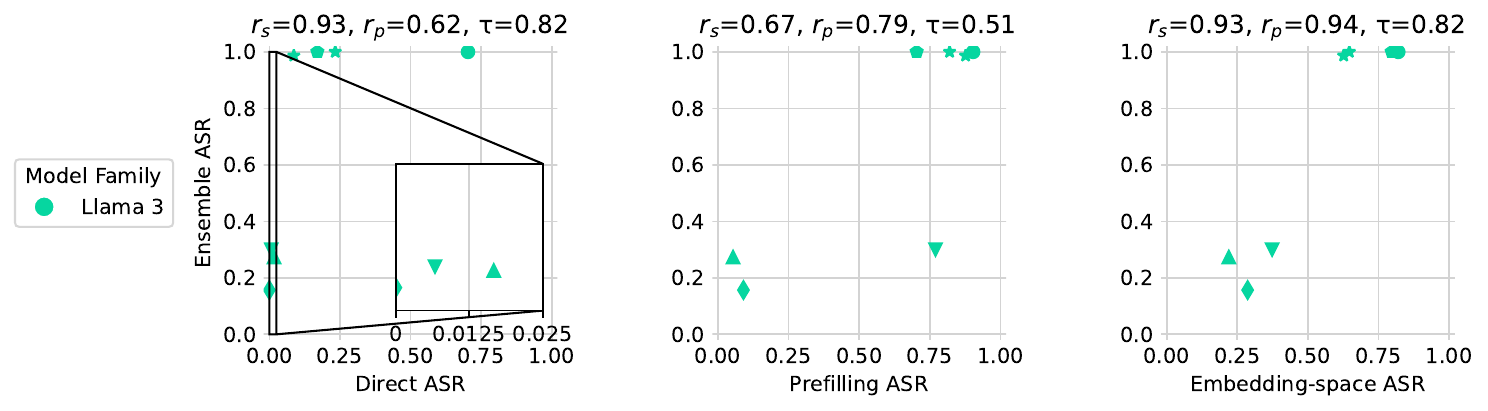}
    \caption{Attack success rates for different variants of Mistral 7B Instruct. We include instruct versions (\instructIcon) as a baseline and compare to safety-tuned (\safetyIcon), adversarially trained (\advIconUp), circuit breaker (\circuitIcon), and capability-optimized (\capabilityIcon) models.}
\end{figure}

These model families were selected due to their popularity and resulting large number of versions.

\subsection{Continuous Adversarial Training}

\begin{figure}[h]
    \includegraphics[width=\textwidth]{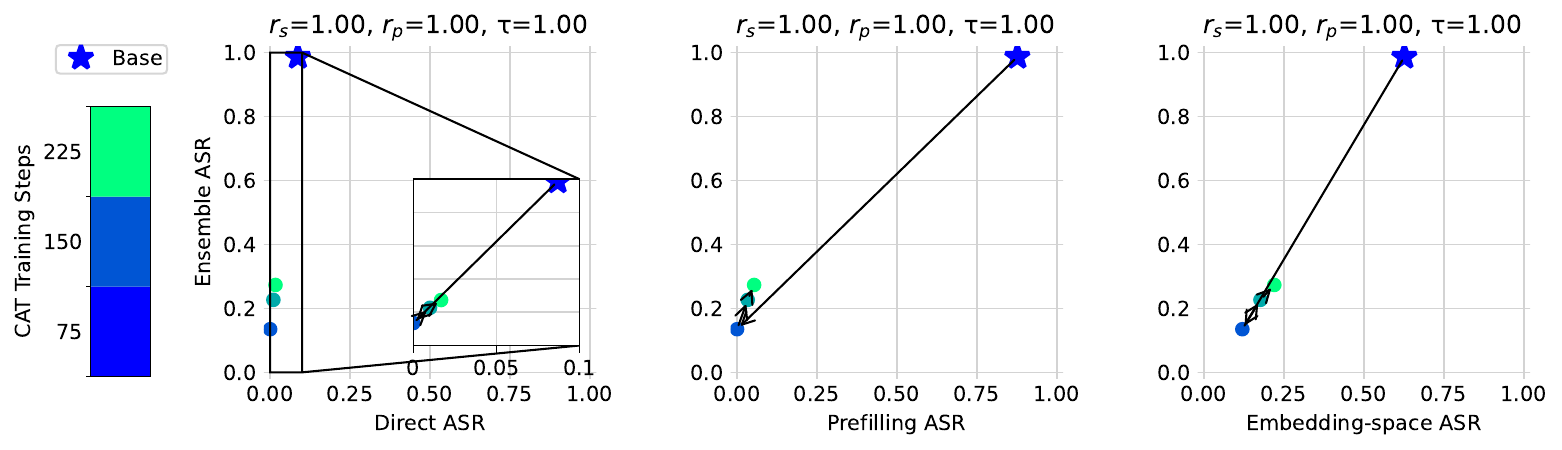}
    \caption{Attack success rates for different number of robustness fine-tuning steps using Continuous Adversarial Training \citep{xhonneux2024efficient} on Llama 3 8B Instruct. All methods are highly correlated with the synthetic red-teamer. Due to resource and time constraints we only compare four training checkpoints. Arrows indicate training progress.}
    \label{fig:combined_correlations_training_capo}
\end{figure}

\end{document}